\documentclass[11pt]{article}

\usepackage{subfigure}
\usepackage{amsmath}
\usepackage{amsthm}
\usepackage{fullpage}
\usepackage{xspace}
\usepackage{url}
\usepackage{graphicx} 

\theoremstyle{remark}
\newtheorem*{example}{Example}

\newcommand{\lm}{LMA\xspace}
\newcommand{\ou}{Ornstein-Uhlbeck\xspace}

\begin{document}

\title{Computational LPPL Fit to Financial Bubbles}

\author{
Vincenzo Liberatore 
\thanks{Division of Computer Science,
Case Western Reserve University,
10900 Euclid Avenue,
Cleveland, Ohio 44106-7071, USA.
E-mail: {\tt vl@case.edu\/}.
URL: {\tt http://vincenzo.liberatore.org/\/}. }
}
\date{}

\maketitle

\begin{abstract}
The log-periodic power law (LPPL) is a model of asset prices during 
endogenous bubbles. 
If the on-going development of a bubble is suspected,
asset prices can be fit numerically to the LPPL law.
The best solutions can then indicate whether a bubble is in progress
and, if so, the bubble critical time (i.e., when the bubble is expected
to burst).
Consequently, the LPPL model is useful only if the data can be fit to the
model with algorithms that are accurate and computationally efficient.

In this paper, we address primarily the computational efficiency and
secondarily the precision of the LPPL non-linear least-square fit.
Specifically, we present a parallel Levenberg-Marquardt
algorithm (\lm) for LPPL least-square fit 
that sped up computation of more than a factor of four over a 
sequential \lm on historical and synthetic price series.
Additionally, we isolate
a linear sub-structure of the LPPL least-square fit that
can be paired with an exact computation of the Jacobian,
give new settings for the Levenberg-Marquardt damping factor,
and describe a heuristic method to choose initial solutions.
\end{abstract}

\section{Introduction}
Financial bubbles and busts have devastating effect on the economy
and on markets.
However, the existence and characteristics of bubbles are notoriously hard
to ascertain if not with hindsight. This paper contributes to the 
investigation of financial bubbles 
within the LPPL framework \cite{lppl-book}, and 
specifically improves on its computational efficiency.

Financial bubbles are hard to detect. In the words of the former
Federal Reserve chairman Alan Greenspan:
``We, at the Federal
Reserve recognized that, despite our suspicions, it was very difficult to 
definitively identify a bubble until after the fact,
that is, when its bursting confirmed its existence.'' \cite{fed-bubble}.
Yet, if a regulatory agency, such as the Fed, were aware of a developing
financial bubble, it could take mitigating steps in the areas of 
monetary policy, 
unconventional open market operations (e.g., \cite{quantitative-easing}),
or supervisory activities (e.g., \cite{imf-macro}). 
Bubble detection is not only critical to the macro-economic decision makers
but it is also helpful to the trader.
In particular, a trading system may requires detailed knowledge of 
the time or the level at which the bubble will burst.

Endogenous financial bubbles have been modeled as a log-periodic power law
(LPPL) \cite{lppl-book}. The LPPL model has two main characteristics:
\begin{itemize}
\item Super-exponential growth, leading to a critical time at which
the asset price will burst (power law), and
\item Oscillations that become progressively faster as the critical
time approaches (log-periodicity).
\end{itemize} 
Super-exponential growth is a sign that price growth are unsustainable.
The oscillatory behavior indicates an incipient system failure,
and is often associated with increasingly more rapid and pronounced swings 
in investor sentiment.
The main issue in LPPL is the fit of the LPPL function to price time series.
Indeed, the LPPL model may be correct, but if it cannot be fit satisfactorily
to price data, its usefulness is limited.
In this paper, we address the computational efficiency and precision 
of the LPPL least-square fit, and present:
\begin{itemize}
\item A parallel algorithm for LPPL least-square fit, 
with a justification and analysis of the speed-up afforded by parallelism.
\item A linear sub-structure of the LPPL least-square fit that
can be paired with an exact computation of the LPPL Jacobian.
\item En route to the algorithm development, we also give new
settings for the Levenberg-Marquardt damping factor and a 
heuristic method to choose initial solutions.
\item An analysis of historical and synthetic price time series
in terms of both computational speed and accuracy.
\end{itemize}

Previous work has paid considerable attention to fitting the LPPL law
to historical time series of financial bubbles, and
a recent summary reviews the state of the art
\cite{lppl-fit}.
Current methods for LPPL bubble detection are currently being
tested in an on-line experiment \cite{sornette-gamble}.
Previous work mostly focus on the issues of accuracy 
and of noise. 
Noise can complicate the estimation of parameters and even make it
hard to distinguish between LPPL (bubble) and exponential (non-bubble)
growth \cite{F01a, CF06a}.
However, if noise has a mean-reverting property, then LPPL can be explain 
many historical bubbles \cite{OU}.
No previous paper has investigated the issue of computational efficiency and 
parallel algorithms specifically for LPPL least-squares.

The paper is organized as follows.
Section \ref{sec:lppl} introduces the LPPL model and
defines the weighted LPPL least-square problem.
Section \ref{sec:init} gives an heuristic method to set multiple 
initial solutions to the LPPL least-square problem.
Section \ref{sec:parallel} describes a parallel implementation of the 
Levenberg-Marquardt algorithm
and Section \ref{sec:damping} gives a method to dynamically set the algorithm's
damping factor.
Section \ref{sec:linear} describes a method to alternate the solution of 
the non-linear least-square problem with the solution of a linear sub-system
while ensuring the exact computation of the Jacobian.
Section \ref{sec:evaluation} describes the evaluation methodology and 
reports on the evaluation results.
Section \ref{sec:conclusion} concludes the paper.

\section{Log-Periodic Power Law (LPPL) and Least Squares}
\label{sec:lppl}
\begin{figure}
\begin{center}
\includegraphics[width=8cm]{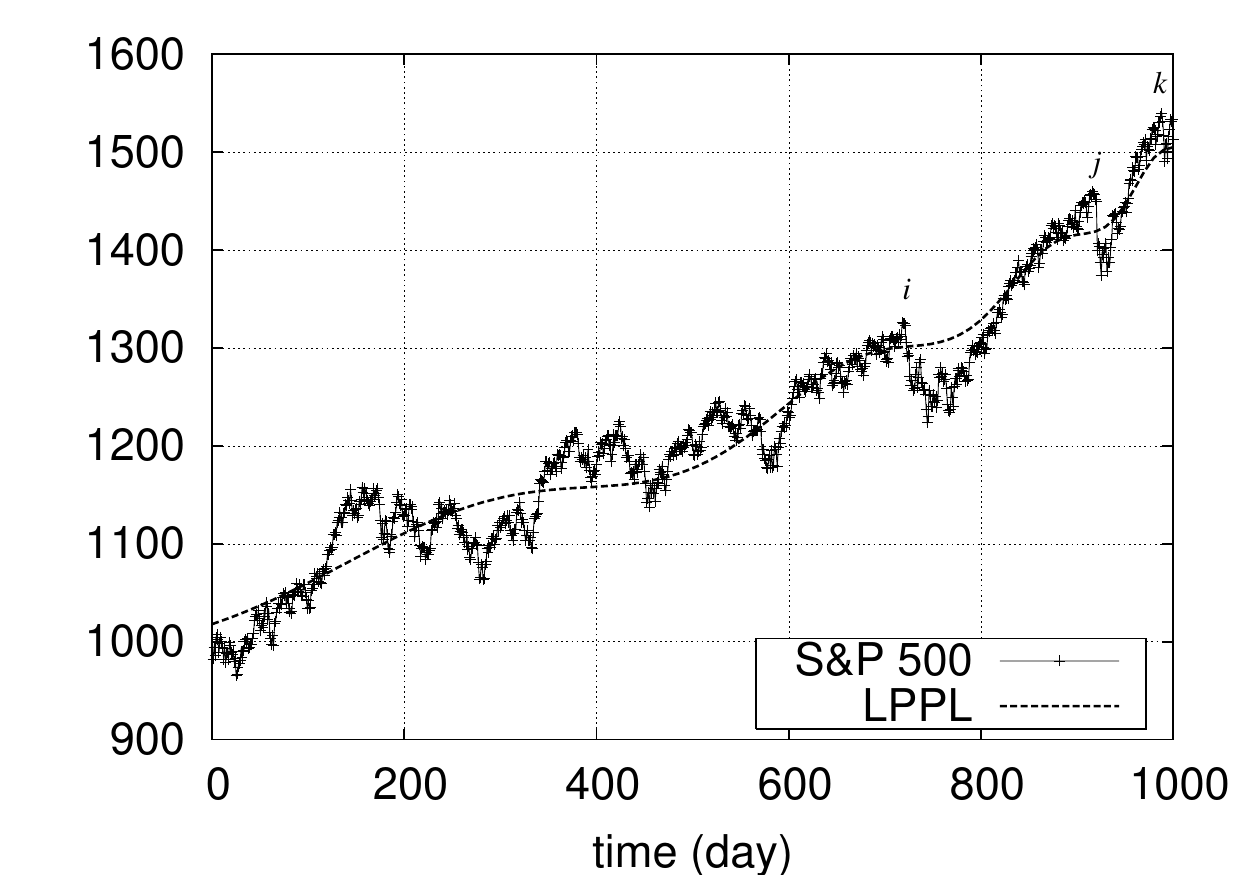}
\end{center}
\caption{S\&P 500 and an LPPL fit from July 2003 to June 2007.}
\label{fig:lppl}
\end{figure}

The {\em log-periodic power law\/} is a function:
\begin{equation}
\label{eq:lppl}
f(x) = A - B (T - x)^m (1 + C \cos (\omega \ln (T - x) + \phi))\;,
\end{equation}
where $B > 0$ and $0 < m \leq 1$. 
The LPPL function is a model for a time series of prices 
$p(1), p(2), \dots , p(n)$ so that $f(i) \simeq \ln p(i)$.
Figure \ref{fig:lppl} shows the S\&P 500 daily closing prices and an 
LPPL fit for the four years between July 2003 and June 2007.
If $m = 1, C = 0$, then LPPL reduces to an exponential fit
of the price time series.
If $m < 1$, then LPPL grows super-exponentially until the
{\em critical time\/} $T$.
If $C, \omega > 0$, then LPPL exhibits oscillations that 
become progressively more frequent as $x$ approaches the
critical time.

Given a time series of prices $p(1), p(2), \dots , p(n)$ and 
weights $w(1), w(2), \dots, w(n)$, 
the {\em LPPL weighted least-square\/} problem is
to find the values of the LPPL parameters
$A, B, T, m, C, \omega, \phi$ ($B > 0, T > n, 0 < m \leq 1$) that minimize
the weighted sum of squared residuals $E = \sum_{i = 1}^n w(i) r_i^2$, where
$r_i = f(i) - \ln p(i) = 
A - B (T - i)^m (1 + C \cos (\omega \ln(T - i) + \phi)) - \ln p(i)$.
We will refer to $E$ as the {\em error\/} corresponding to the given
LPPL parameters, and 
to $E / d$ as the {\em average error\/},
where $d = n - 7$ is the {\em degrees of freedom\/} of the 
least-squares problem.

The LPPL weighted least-squares can be solved with various methods,
such as the {\em Levenberg-Marquardt algorithm (\lm)\/} \cite{levmar}.
The \lm is an iterative method that maintains a vector of parameters
$P = (A, B, T, m, C, \omega, \phi)$ and at each iteration adjusts
$P$ so as to reduce the error $E$. 

Previous work has considered unweighted LPPL least-squares ($w(i) = 1$) on 
sub-intervals $[s, t] \subseteq [1, n]$ \cite{lppl-fit}. 
Weighted least-squares generalizes sub-interval fitting since it 
reduces to sub-interval fitting in the special case
\begin{equation}
\label{eq:step-weight}
w(i) =  \begin{cases}
0 & \text{if $i \not\in [s, t]$} \\
1 & \text{if $i \in [s, t]$}
\end{cases}
\end{equation}
Weighted least-squares make it possible to emphasize recent 
history over older data.
For example, if weights are as in \eqref{eq:step-weight} and $t = n$, then 
the fit will only depend on the recent interval $[s, n]$.
As another example, if 
\begin{equation}
\label{eq:quad-weight}
w(i) = \left(\dfrac{W}{n - i + W}\right)^2\;,
\end{equation}
the importance of $p(i)$ decreases quadratically with its age $n - i$.
The importance of recency can be increased by taking smaller values of $W$.
Returning to the case of general weights, 
recency can be progressively emphasized by decreasing weights more rapidly 
for progressively older data points.

Recency weighting is relatively less important during the last stages of 
a bubble. 
When a bubble is approaching its critical time $T$, 
recent prices are much larger than older prices.
Hence, the residual $r_i$ is much larger for recent data than for older prices,
even if the relative error $\ln p(i) / f(i)$ is the same.
Thus, the least-square objective is affected more by recent data points than older 
ones.
However, recency weighting is important when the bubble is in its early 
stages because price values are relatively close to each other.

\section{Initial Solutions}
\label{sec:init}
Non-linear least-squares is in general a non-convex optimization problem, and
a solution algorithm, such as \lm, can reach a local minimum 
that is not globally optimum.
The existence of local minima can be partly obviated by starting \lm from
multiple initial solutions.
Initial solutions can be selected via Taboo search \cite{lppl-fit}.
Yet, in the presence of a strong LPPL component to price data, 
initial solutions can also be chosen by inspection of the price series.

In this paper, the initial solutions are determined as follows. 
First, we find the best fit of prices to an exponential function of
the form $\ln p(i) = A - B (T - i)$.
The exponential fit is found by linear least-squares, and
can be regarded as fitting the LPPL function \eqref{eq:lppl}
under the additional constraints that $m = 1$ and $C = 0$.
The \lm method starts with these exponentially fit parameters and moves on
to a general LPPL fit, where the constraints $m = 1, C = 0$ can be 
violated.
However, if the \lm general LPPL fit does not significantly reduce the error,
it can be taken as an indication that no LPPL bubble is forming.
In other words, the exponential starting point can be interpreted informally 
as testing for the null hypothesis that prices grow exponentially, 
i.e., no super-exponential LPPL bubble is developing.

Second, we estimate the initial values of $\omega$ and $T$ from price 
gyrations.
We say that two indexes $i$ and $j > i$ are {\em consecutive in angle\/} if 
$\omega \ln (T - i) = \omega \ln (T - j) + 2 \pi$.
The estimation of $\omega$ and $T$ relies on the fact that 
in many cases it is easy to identify visually two indexes
that are consecutive in angle.
For example, two consecutive peaks of the LPPL function are
consecutive in angle (points $j$ and $k$ in Fig. \ref{fig:lppl}).
If two indexes $i < j$ are consecutive in angle, 
then $(T - i) / (T - j) = e^{(2 \pi) / \omega}$.
Define $\rho = e^{(2 \pi) / \omega}$.
Then, $\omega = (2 \pi) / \ln \rho$ and $T = (\rho j - i) / (\rho - 1)$.
To estimate $\rho$, consider three points $i < j < k$ such that $i, j$ 
are consecutive in angle and $j, k$ are consecutive in angle.
Then, $j - i = \rho^l (\rho -1)$ for some integer $l$, and
$k - j = \rho^{l-1} (\rho - 1)$, so that 
$\rho = (j - i) / (k - j)$.

An estimate $i$, $j$, and $k$ is obtained by selecting three consecutive
price peaks (or troughs). 
It is an open question to automate the selection of $i, j, k$,
for example, with a visual pattern recognition algorithm.
At any rate, each choice of $i, j, k$ corresponds to a value of $\rho$,
and thus to a different initial solution.
The \lm process will be started from multiple initial parameters.

It remains to choose an initial value of $\phi$.
We use the previously chosen initial values of $T$ and $\omega$,
and without loss of generality make the assumption that $C \geq 0$.
If $i$, $j$, and $k$ are chosen as approximate price peaks,
then $\omega \ln (T - k) + \phi = \pi$, 
and so $\phi =  \pi - \omega \ln (T - k)$.
Similarly, the LPPL equation \eqref{eq:lppl} can be solved for $\phi$
if $i$, $j$, and $k$ are chosen as approximate price troughs, or at any other 
fixed angle.

\begin{example}
In the 2003-2007 S\&P 500 bubble up to 6/30/2007 (Fig. \ref{fig:lppl}),
three consecutive peaks are $i = 718$ (5/5/2006), $j = 916$ (2/20/2007), and 
$k = 988$ (6/4/2007).
Hence, $\rho = (916 - 718) / (988 - 718) = 2.75$, 
$T = (\rho k - j) / (\rho - 1) = 1029.14$,
$\omega = 2 \pi / \ln (\rho) = 6.21$, and 
$\phi = -19.95$.
The initial linear least-squares fit to an exponential function leads to
$A = 9.46$ and $B = 3.53 \cdot 10^{-4}$, with an error of 
$0.9$.
The initial point is subsequently fit to LPPL,
and the final value of the parameters is 
$A = 7.47, B = 0.014, T = 1105.08$ (11/16/2007), 
$m = 0.52, C = -0.04, w = 9.87, \phi = 2.72$, 
with an error of $0.65$.
The final parameter values and the substantial error reduction denote
that a bubble was likely in progress.
The actual S\&P 500 peak was on 10/9/2007.
\end{example}

\section{Parallel Least-Squares}
\label{sec:parallel}
Modern computer systems provide significant parallelization capabilities:
most CPUs have multiple cores, multiple CPUs and co-processors are 
placed on the same motherboard, general-purpose GPUs 
execute streams of independent operations on multiple data,
and cloud computing makes it possible to gain 
inexpensive access to a vast number of computational resources. 
Furthermore, the amount of parallelism is expected to increase in the next 
few years as current trends solidify: more cores will be placed on a single CPU,
and cloud computing will gradually become commonplace.
Several programming interfaces (e.g., OpenMP \cite{openmp}) make it easy to 
develop parallel programs, especially if the computation involves primarily 
iterations of independent statements. 
Previous work describes a general-purpose parallel \lm implementation based
on three techniques:
solving multiple problems in parallel, 
parallelizing the computation of the Jacobian, and 
probing multiple values of the damping factor in parallel \cite{parallel-lm}. 
Parallelism offers an opportunity to implement faster \lm especially tailored
to the LPPL least-squares.

Parallelism can be exploited at various levels of granularity. 
At the coarsest level, multiple securities can be analyzed in parallel.
Moreover, each security can be analyzed from multiple initial solutions and
for different weight functions.
Since different securities, initial solutions, and weights lead to independent
\lm executions, the analysis can be conducted in parallel.
Widespread \lm tools, such as levmar \cite{levmar}, are thread-unsafe, and 
a single process cannot accommodate concurrent \lm invocations.
Hence, multiple securities, weights, and starting points are analyzed by 
independent processes. 

At the other extreme, fine-grained parallelism
can be used to speed-up the computation of the LPPL function \eqref{eq:lppl}
and associated quantities.
Each \lm iteration requires the computation of $f$ and 
of its Jacobian $J$ at the current solution $P$.
In general, the Jacobian $J$ can be computed either analytically or through
approximation methods, such as finite differences.
Since $f$ has a relatively simple closed expression, 
we compute the LPPL Jacobian $J$ analytically. 
In practice, each step requires the computation of $f$ at $n$ points and 
the evaluation of 7 partial derivatives at the same $n$ points. 
The computation at $i$ is independent of the
computation at $j \neq i$, and so its parallelization is 
a perfect match for the OpenMP ``parallel for'' construct.
The computation of $f$ and $J$ can be parallelized even 
if the \lm solver is not thread-safe  
because it is invoked from within the solver.
In this case, the program is sequential except at the very bottom of the
invocation tree, where $f$ and $J$ are calculated by multiple threads.

\begin{table}
\begin{center}
\begin{tabular}{|r|l|} \hline
{\bf Time percentage\/} & {\bf Function Name\/} \\ \hline
19.0321 &  cos \\
17.3662 &  \_\_ieee754\_log \\
16.1608 &  \_\_ieee754\_pow \\
10.7317 &  lppl\_vector.omp\_fn.2 \\
10.4502 &  \_\_exp1 \\
 2.5473 &  pow \\
 1.2532 &  isnan \\
 1.0655 &  log \\
 1.0591 &  finite \\
 0.6072 &  \_\_mul \\
 0.4087 &  dlevmar\_L2nrmxmy \\
 0.3768 &  /usr/lib64/libblas.so.3.0.3 \\
 0.3049 &  levmar\_weight \\
 0.2072 &  sin \\
 0.1884 &  csloww1 \\
 0.1285 &  \_int\_malloc \\ \hline
\end{tabular}
\end{center}
\caption{Function execution time collected with oprofile during a sample run.
Numerical functions (e.g., power, cosine, logarithm) appear only in the 
computation of the LPPL function \eqref{eq:lppl} and its Jacobian.}
\label{table:oprofile}
\end{table}
The computation of $f$ and $J$ accounts for most of the program execution time. 
Table \ref{table:oprofile} shows a breakdown of the running time: the largest
contribution comes from numerical functions 
(such as raising to power, cosine, logarithm) 
that appear solely in the evaluation of the LPPL function and of
its Jacobian.
In summary, the parallelization of $f$ and $J$ is not only relatively easy, 
but it is also  likely to lead to a large performance improvement.
On the other hand, it is not clear {\it a priori\/} that multiple threads
speed execution up since the running time depends also on the work overhead needed
to manage a thread pool.

The parallel computation of the LPPL function and of its Jacobian
requires a parameter that sets the number
of concurrent execution threads. 
The thread pool size depends on the number
of cores available to the process.
Consequently, the available core count depends 
not only on the underlying computational resources, 
but also on the number of securities, starting points, and weight
functions that are simultaneously running.

\section{Damping Term}
\label{sec:damping}
The \lm algorithm maintains a {\em damping term\/} $\mu$,
which represents the contribution of the steepest descend
method to the next iteration.
If $\mu$ is large, then \lm is approximately steepest descent, 
and if $\mu = 0$, then \lm reduces to a linear least-square applied to 
the linearization of the fit function around the current
solution.
The \lm algorithm dynamically adjusts the value of $\mu$, 
it always keeps $\mu > 0$, 
and it can increment or decrement $\mu$ at any iteration by
multiplying its current value by certain carefully
selected factors.
The \lm algorithm can terminate with fit parameters $P$
because of certain circumstances 
(e.g., singular matrices) in which 
the algorithm should be restarted from the current solution
$P$ with a larger value of $\mu$.
Furthermore, it is possible that the algorithm reaches
$\mu = 0$ because of underflow, making it impossible to 
subsequently increase $\mu$ by multiplying it by any factor.
If \lm terminates with $\mu = 0$ or
in the other cases that require restarting from
a larger value of $\mu$, 
then we double the value of a parameter $\bar{\mu}$ and 
restart \lm with the initial value of $\mu$ set to 
$\bar{\mu}$.
The $\bar{\mu}$ parameter is also subject to a global upper bound
beyond which it cannot be further increased.

\section{Linear Sub-System}
\label{sec:linear}
If $T$, $m$, $\omega$, $\phi$ are constrained to fixed values,
then the least-square problem on $A$, $B$, and $C$ is linear.
Linear least-squares are typically
fast since they require only the computation of one exact
Jacobian and the solution of a system of linear equations.
However, computation speed must be weighed against solution error.
In general, constraining parameters will lead to a solution with
a larger error than the unconstrained global optimum.
However, the \lm algorithm can fall into a local minimum,
and the relation between its final parameters and
the linear least-square is not always clear.
In particular, when the non-linear \lm terminates at a solution $P_1$,
if the non-linear parameters $T$, $m$, $\omega$, $\phi$
are fixed at the $P_1$ value,
the linear least-square on $A$, $B$, and $C$ often leads
to a solution $P_2$ that improves on $P_1$, i.e., its error is smaller. 
Furthermore, if the non-linear \lm is restarted from $P_2$, 
the \lm may be able to further improve on $P_2$,
whereas \lm had not been able to make progress from $P_1$.
In some sense, the linearized system can be viewed as
a computationally efficient way to extricate \lm from a local minimum.
Alternatively, previous work derived analytical expressions for 
$A, B, C$ as a function of the other parameters by expanding the 
linear least-squares optimality conditions.
These expressions for $A, B, C$ are then substituted in the 
non-linear least-square problem \cite{lppl-fit}.
Direct substitution reduces the number of parameters in the fit, 
but it complicates the LPPL expression and especially the 
analytical calculation of the Jacobian matrix. 

To interleave \lm with linear least-squares,
we run \lm until either it terminates or it has reached the $L$th iteration,
and then solve the linear least-square assuming that 
$T$, $m$, $\omega$, $\phi$ are fixed. 
If the linear least-square improves the error,
it is taken as the starting point for another round of the \lm algorithm. 
The process terminates when the linear least-square fails to improve the 
current solution and the damping factor $\mu$ cannot be further increased.

The parameter $L \geq 1$ is a bound on the number of \lm iterations 
and can be supplied by the user.
We have also implemented an adaptive method to choose $L$ depending on the 
relative gains made by the linear and the non-linear algorithm.
The adaptive method changes the value of $L$ depending on 
which least-square algorithm reduces the error the most during 
a unit of computation time. 
In other words, the adaptive method will run \lm longer if \lm reduces
the error more than linear least-squares during a unit of computation time,
and will run the linear least-square more often if conversely the linear
least-square reduces the error more than \lm during a unit of 
computation time.
The method collects the error reduction and 
the running time of each invocation of both linear least-squares and LMA. 
In the case of linear least-squares,
we estimate the error reduction per unit of time
by simply dividing the measured error reduction by
the execution time.
The \lm unit error reduction is more complicated to estimate because
\lm executes various initializations regardless of how many iterations 
the algorithm executes. 
The \lm run time is modeled as $T = T_1 \ell + T_0$, where $T_0$ is 
the time needed to execute the initialization, $\ell \leq L$ is the
number of iterations, and $T_1$ is the time needed
for each iteration. 
The values of $T_0$ and $T_1$ are adaptively estimated
by keeping track of the execution time during the last 
two invocation of \lm with different values of $\ell$.
We then compute an estimate of the 
error reduction that would occur if the \lm computation
took one more unit of time.

Given the linear and non-linear estimates of error reduction per unit of time,
we compare them to choose whether to increase or 
decrease $L$. The adaptive mechanism proceeds in two phases:
a start-up phase and a regime phase.
The start-up phase aims at finding quickly an approximate value for
the best value of $L$, whereas the regime phase aims toward 
a more precise value of $L$.
The start-up begins with a small user-supplied value of $L$,
and doubles it until the marginal unit \lm error improvement
is smaller than the unit error improvement of the linear least-squares.
The regime phase increases or decreases $L$ by one
depending on which algorithm leads to the largest unit error
reduction.

\section{Evaluation}
\label{sec:evaluation}

\subsection{Methodology}
The LPPL least-square fit can be carried out under two main
goals: prediction of a bubble prior to its critical time ($n < T$),
or fit of asset prices to LPPL after the fact to verify whether 
LPPL models the evolution of prices ($n = T$).
The focus of this paper is on prediction ($n < T$).

The evaluation uses real and synthetic data sets.
Real data sets are daily closing values of major securities and 
stock indexes described in Table \ref{table:dataset}.
\begin{table}
\begin{center}
\begin{tabular}{|l|l|l|} \hline
{\bf Series\/} & {\bf Period\/} & {\bf Data Points\/} \\ \hline\hline
Dow Jones & June 1921-July 1929 & 2440 \\ \hline
S\&P 500  & July 1985--July 1987 & 527 \\ \hline
S\&P 500  & July 2003--June 2007     & 1000 \\ \hline
GLD       & March 2009--October 2009 & 171 \\ \hline
\end{tabular}
\end{center}
\caption{Real price time series used in the evaluation.}
\label{table:dataset}
\end{table}

Synthetic data sets are generated by adding noise
to LPPL price values:
$\log p(i) = f(i) + \sigma B(i)$,
where $B(i)$ is the value of a Brownian motion process at time $i$ and 
$\sigma$ is a scaling factor \cite{JSL99a, JLS00a}.
Geometric Brownian motion is commonly used to model
price dynamics in an efficient market (e.g., \cite{black-scholes}),
where $\sigma$ is called {\em implied volatility\/}.
Thus, synthetic data can be regarded the combination
of a super-exponential bubble with an efficient market.
Recent work proposes a more sophisticated model based
on LPPL plus both Brownian motion and an \ou process \cite{OU}.
However, Brownian motion is the most difficult
noise component for the LPPL least-square fit
\cite{OU, F01a, CF06a}.
Synthetic data were generated with the parameters shown in 
table \ref{table:synthetic}.
\begin{table}
\begin{center}
\begin{tabular}{|l|r|r|r|} \hline 
& {\bf Base\/} & {\bf Oscillatory\/} & {\bf Exponential\/}
\\ \hline\hline
$A$ & 5 & 5 & 5 \\ \hline
$B$ & 0.02 & 0.02 & 0.005\\ \hline
$T$ & 1100 & 1100 & 1100 \\ \hline
$m$ & .68 & .68& 1 \\ \hline
$C$ & .05 & 0.2 & 0 \\ \hline
$\omega$ & 9 & 9 & 1 \\ \hline
$\phi$ & 0 & 0 & 0 \\ \hline
$n$ & 1000 & 1000 & 1000 \\ \hline
$\sigma$ & 0.005 & 0.02 & 0.05 \\ \hline
\end{tabular}
\end{center}
\caption{Parameters used in synthetic trace generation.}
\label{table:synthetic}
\end{table}
Base and oscillatory traces capture a LPPL processes, and exponential traces
capture non-LPPL processes.
The base case contains a log-periodic term ($C, \omega > 0$)
that is relatively small so that the LPPL function is always increasing
over time.
The oscillatory model assumes a larger value of $C$ that magnifies
the log-periodic oscillations.
For each stochastic model (base, oscillatory, exponential), five random
traces were generated.

The computing platform was an 8-core 3.4Ghz Intel Xeon with 8GB RAM
running Linux CentOS (RHEL) 5.
The underlying \lm solver is levmar \cite{levmar}, and the computation
of the LPPL function and its Jacobian employed OpenMP \cite{openmp}.

\subsection{Parallelism}
Each least-square LPPL fit was run on an 8-core machine as a process
with 8 threads of control. The machine was otherwise unloaded
(except for various unavoidable and lightweight system processes),
and the LPPL fit achieved close to 100\% CPU utilization.
The parallel least-square typically took less than 1/4 of the time of the
sequential implementation, i.e., we observe a speed up of a factor of 
four or more. 
Since the parallel least-square has
8-fold parallelism, the amount of work increased significantly. 
In particular, more than 1/3 of the running
time is taken by OpenMP management (Table \ref{table:parallel-oprofile}).
\begin{table}
\begin{center}
\begin{tabular}{|r|l|} \hline
{\bf Time percentage\/} & {\bf Function Name\/} \\ \hline
36.1513 &     /usr/lib64/libgomp.so.1.0.0 \\
12.3288 &     \_\_ieee754\_pow \\
11.6925 &     \_\_ieee754\_log \\
11.3157 &     cos \\
 8.4633 &     lppl\_vector.omp\_fn.2 \\
 6.9043 &     \_\_exp1 \\
 2.9887 &     pow \\
 2.5607 &     isnan \\
 1.5655 &     log \\
 1.1870 &     /no-vmlinux \\
 0.8585 &     dlevmar\_L2nrmxmy \\
 0.6399 &     /usr/bin/oprofiled \\
 0.4110 &     /usr/lib64/libblas.so.3.0.3 \\
 0.3673 &     levmar\_weight \\
 0.3386 &     \_\_mul \\
 0.3371 &     finite \\
 0.2273 &     jac\_lppl\_vector.omp\_fn.1 \\
 0.2170 &     sin \\ \hline
\end{tabular}
\end{center}
\caption{Function execution time during a sample parallel run.
libgomp is the gcc OpenMP v. 3.0 shared support library.}
\label{table:parallel-oprofile}
\end{table}

\paragraph*{Discussion.}
The optimal degree of parallelism depends on the number of jobs and 
on the number of available cores. 
For example, while a single job was than 4 times faster on 8 cores,
eight distinct jobs complete sooner if each is sequential.

The workload is dominated by the independent execution 
of the same set of instructions on multiple data. 
Therefore, it is likely that the computation would benefit from SIMD
(single-instruction multiple data) and related hardware architecture,
such as GPGPU (General-Purpose computation on Graphics Processing Units)
\cite{gpgpu}.
By contrast, general multi-core architectures, such as the one used in this 
paper, are easier to program but may have lower performance than
specialized SIMD.
This paper leaves it as an open problem to implement the LPPL least-square 
on GPGPU.

Another disadvantage of general multi-core architecture
is that the parallelizable data set (the {\em stream\/}) is relatively small.
In practice, a price time series has few thousands points at most, and
corresponds to a full stream. As a result, the overhead of managing multiple
threads over a multi-core architecture is a substantial fraction of the 
execution time.
On the flip side, the algorithm is scalable with the number $n$ of data
points, so the thread management overhead becomes smaller if more data 
is available.

\subsection{LPPL Least-Square Fit}
\paragraph*{Historical Time Series.}
Table \ref{table:result} gives the error reduction of the best least-square
LPPL fit, and 
the predicted and actual critical time in the minimum residual error fit.
As an example, Figure \ref{fig:lppl} gives the best fit of the S\&P 500 (2007)
prices.
The substantial error reduction and the final parameter values are strong
evidence that an endogenous LPPL bubble was in progress.
However, the best fit predicts a critical time that deviates, in some cases 
significantly, from the actual bubble burst.
Critical time overestimation was robust to several choices of a 
quadratic weight function \eqref{eq:quad-weight}.
\begin{table}
\begin{center}
\begin{tabular}{|l|l|l|l|l|l|l|l|} \hline
& \multicolumn{2}{|c|}{\bf Average Error\/} & \multicolumn{2}{|c|}{$T$} & & & 
\\ \hline
{\bf Series\/} & {\bf Exponential fit\/} & 
{\bf LPPL fit\/} & {\bf Fit\/} & {\bf Actual\/} & 
$m$ & $|C|$ & $\omega$ \\ \hline\hline
Dow Jones 1929 & $1.230 \cdot 10^{-2}$ & $0.294 \cdot 10^{-2}$ & 2784 & 2467 & 
0.372 & 0.0414 & 12.23 
\\ \hline
S\&P 500 1987  & $1.817 \cdot 10^{-3}$ & $0.516 \cdot 10^{-3}$ & 658 & 573 & 
0.417 & 0.0705 & 9.89
\\ \hline
S\&P 500 2007  & $9.007 \cdot 10^{-4}$ & $6.472 \cdot 10^{-4}$ & 1105 & 1071 & 
0.518 & 0.0438 & 9.87
\\ \hline 
GLD 2009       & $8.405 \cdot 10^{-4}$ & $3.102 \cdot 10^{-4}$ & 211 & 195 & 
$0.285 \cdot 10^{-3}$ & $7 \cdot 10^{-5}$ & 18.40 
\\ \hline
\end{tabular}
\end{center}
\caption{Error and parameter values of the best fit of historical bubble
price series.}
\label{table:result}
\end{table}

The best LPPL solutions were characterized by a strong local correlation 
between some of the variables. 
In particular, in the neighborhood of the optimum, 
$\omega$ had absolute correlation higher than $0.9$ with both $T$ and $\phi$.
In other words, even if $\omega$ had been fixed to a value slightly
different from the optimum (e.g., \cite{hkfit}), the LPPL fit would
still be able to make progress by tweaking the values of $T$ and $\phi$.

\paragraph*{Synthetic Time Series.}
On the base synthetic traces (Table \ref{table:synthetic}), 
the best fit almost always predicts
larger values of the critical time than in the underlying process.
The overestimation was 63 trading periods on average and ranged 
between -5 and +166 trading periods.
On the oscillatory traces, the best fit almost always predicts
smaller values of the critical time than in the underlying process.
The underestimation was 19 trading periods on average and 
ranged between -42 and +67 trading periods.
We conjecture that in the presence of additive noise such as Brownian motion
the best least-square fit tends to overestimate $T$ if oscillations
are muted ($C, \omega$ small) and 
to underestimate it if oscillations are larger ($C, \omega$ large).
The hypothesis matches the best fit behavior on the historical 
price series (Table \ref{table:result}),
where $C$ is small and comparable to that of the base synthetic trace,
oscillations are muted (for example, see Figure \ref{fig:lppl}), and 
estimated critical times $T$ always exceeded actual critical times.

On the exponential synthetic traces, the fit found either $m \simeq 1$,
which denotes a nearly exponential fit,
or $\omega \simeq 1$, in which case
the log-periodic oscillations were so small that the LPPL fit did 
not significantly differ from a super-exponential function.
Parameter values $m, \omega \simeq 1$ denote poor statistical
significance for the LPPL fit \cite{misfit}.

\paragraph*{Discussion.}
The least-square LPPL algorithm distinguished clearly between LPPL and
non-LPPL growth. In particular, it was sufficient to check for
$m \simeq 1$ and $\omega$ small (say, $\omega \simeq 1$).
The implication is that it is possible for, say, a regulatory
agency, to distinguish between endogenous bubbles and ordinary 
exponential growth.
However, the estimated parameters differed from their true value,
and in particular the critical time $T$ is underestimated
in the presence of Brownian noise and if the underlying LPPL
function lacks strong oscillations.
As a consequence, the best LPPL fit was not a workable indicator 
for technical analysis. 


\section{Conclusion}
\label{sec:conclusion}
In this paper, we have presented methods for the solution of the 
LPPL weighted least-square fit via the Levenberg-Marquardt algorithm. 
The method heavily relies on parallelism to 
cut down on the critical computation path, namely the evaluation of
the LPPL function and of its Jacobian.
We have described cases that are based on historical price series
and where the method cuts the solution time by a factor of four or more.
We have also presented a method to dovetail the non-linear LPPL least-square
with the solution of the linear LPPL sub-component while preserving the 
exact computation of the Jacobian.
En route, we have described methods for selecting initial solutions
to the LPPL fit and a method for setting the \lm damping factor. 

The paper leaves a number of open problems.
On the computational side, the running time is dominated by the 
stream computation of the LPPL function and of its Jacobian.
When this computation is parallelized on a general-purpose multi-core
process, the thread management overhead approached 40\% of the 
process work. 
Thus, much promise is held by alternative architectures, such as SIMD
or GPGPU.
On the economics side, the method is a tool for further research on LPPL, and 
especially for deriving precise fits of LPPL to noisy price data.

\bibliographystyle{plain}
\bibliography{lppl}

\end{document}